\documentclass[11pt,a4paper]{article}
\usepackage[T1]{fontenc}
\usepackage{amsmath,amssymb,amsthm}
\usepackage[a4paper,margin=24mm]{geometry}
\usepackage{titlesec,xcolor}
\usepackage{enumitem}
\usepackage{cite}
\usepackage[hidelinks]{hyperref}
\usepackage[nameinlink,noabbrev,capitalize]{cleveref}
\usepackage{graphicx}

\titleformat{\section}{\normalfont\large\scshape\centering}{\thesection.}{0.75em}{}
\titleformat{\subsection}[runin]{\normalfont\normalsize\bfseries}{\thesubsection.}{0.5em}{}[.]
\titleformat{\subsubsection}[runin]{\normalfont\normalsize\itshape}{\thesubsubsection.}{0.5em}{}[.]

\theoremstyle{plain}
\newtheorem{theorem}{Theorem}[section]
\newtheorem{proposition}[theorem]{Proposition}
\newtheorem{lemma}[theorem]{Lemma}

\theoremstyle{definition}
\newtheorem{definition}[theorem]{Definition}

\theoremstyle{remark}
\newtheorem{remark}[theorem]{Remark}
\numberwithin{equation}{section}

\newcommand{\R}{\mathbb{R}}

\newcommand{\Hplus}{\mathbb{H}_+}

\newcommand{\BG}{\mathcal B_\Gamma}

\newcommand{\Uhat}{\widehat U}
\newcommand{\Utilde}{\widetilde U}
\newcommand{\dd}{\,\mathrm d}
\newcommand{\ii}{\mathrm i}

\renewcommand{\Re}{\operatorname{Re}}
\renewcommand{\Im}{\operatorname{Im}}
\DeclareMathOperator{\Res}{Res}
\DeclareMathOperator{\Tr}{Tr}

\usepackage{authblk}

\allowdisplaybreaks[1]

\title{Resurgent Analysis of the Single-Orbit Term in the Selberg Trace Formula}
\author{Wen Shen}
\affil{\small Department of Mathematics, Capital Normal University, Beijing 100048, P. R. China}
\date{}

\begin{document}
\maketitle
\begin{abstract}
We study the large-time asymptotic expansion, the Borel-Laplace transforms and the Stokes phenomenon of the single-orbit term in the Selberg trace formula. By a single-orbit term we mean the contribution from a primitive closed geodesic orbit and its iterations on a compact hyperbolic surface, and its formal large-time expansion is divergent. We prove that the divergent formal series is uniformly $1$-Gevrey asymptotic to the convergent single-orbit term and recover the single-orbit term by a Laplace integral along the positive real axis. Other directions for the Laplace integration are also explored by analyzing the singularities of the Borel transform of this $1$-Gevrey formal series. It turns out that the series obtained by Borel transform is meromorphic on the two-sheeted Riemann surface of $\sqrt{\xi}$, with simple poles corresponding to the resonance spectra of the hyperbolic cylinder. The Laplace transform can be performed along almost every direction on the two-sheeted Riemann surface. Finally, the Stokes structure as the limit of the difference between Laplace integrals along different directions can be computed by a convergent sum of the exponential of resonance spectrum. These results demonstrate the phenomenon that the resurgent structure of a single-orbit term in the Selberg trace formula is reflected by its associated cylinder resonance spectra.
\end{abstract}


\section{Introduction}\label{sec:introduction}
\subsection{Background and Motivation}
The relation between quantum spectra and classical dynamics is a central theme in semiclassical physics. The seminal Gutzwiller trace formula \cite{Gutzwiller1971,Gutzwiller1980} describes the spectral density through classical periodic orbits asymptotically. The well-known Selberg trace formula \cite{1972Mckean} relates the spectra of the Laplace operator on a compact hyperbolic surface to the lengths of its closed geodesics as classical free motion in an exact form. Plugging the heat kernel as test function into it, there is an exact relation between the spectral sum and the sum of geodesic orbit terms. This raises a natural question that how spectral information emerged in the asymptotic expansion of these orbit terms? Can the spectrum be visible in a series about a primary geodesic orbit and its iterations?

Borel-Laplace resummation in resurgence theory \cite{SauzinIntro,Sauzin} provides a possible approach to such questions. It relates a factorially divergent formal asymptotic expansion to a locally convergent series by Borel transform. Its analytic continuation and the singularities
in the Borel plane contain valuable information not visible in any finite truncation of the formal divergent expansion. The singularities of the analytic continuation determine the difference between different Borel-Laplace summations, which is also interpreted as "the Stokes phenomenon".

The connection between Selberg trace formula and resurgence has a substantial history. Cartier and Voros~\cite{CartierVoros} developed a complex-analytic interpretation of the Selberg trace formula, and Voros~\cite{Voros} related the singular points of the Borel transform of a spectral function to the geodesic lines in Selberg trace formula. Dunne~\cite{Dunne} analyzed the different Borel structures of short- and long-time heat-kernel expansions on hyperbolic space. More recently, with Sun we~\cite{ShenSun} used a theta function defined through square-root spectral parameters to study regularized determinants by resurgence theory.

Different from the objects previously studied in the trace formula, we take the Borel-Laplace summation on the long time asymptotic expansion of a primitive orbit sum, and relate the singularities on the Borel plane to the hyperbolic resonance spectra. Then we can characterize the Stokes phenomenon for the long-time asymptotic expansion of orbit terms via the associated spectra.

\subsection{Selberg trace formula and Main Results}

Let $X$ be a connected compact oriented hyperbolic surface of curvature $-1$, and $\Delta_X$ its Laplace-Beltrami operator. Denote by $\mathcal P_X$ the set of primitive oriented closed geodesic orbits on $X$, and by $\ell_\gamma$ the length of $\gamma\in\mathcal P_X$. The Selberg trace formula \cite[section 3.2]{1972Mckean} says that,
\begin{equation*}
 \Tr e^{-\tau(\Delta_X-1/4)}
 =\frac{\operatorname{Area}(X)}{4\pi}\int_{\mathbb R}r\tanh(\pi r)e^{-\tau r^2}\dd r+\sum_{\gamma\in\mathcal P_X} U_{\ell_\gamma}(\tau),
\end{equation*}
for all $\tau\in\Hplus:=\{\tau\in\mathbb C:\Re\tau>0\}$, where $U_{l_\gamma}(\tau)$ is the sum of a convergent series on $\Hplus$. It is developed by a primitive oriented closed geodesic $\gamma$ and all its iterations
\begin{equation}\label{eq:objects}
U_{\ell_\gamma}(\tau):=\frac{1}{\sqrt{4\pi\tau}}\sum_{k\geq1}b_k(\ell_\gamma)e^{-(k\ell_\gamma)^2/(4\tau)},
\text{ and }b_k(\ell_\gamma):=\frac{\ell_\gamma}{2\sinh(k\ell_\gamma/2)}.
\end{equation}
We call $U_{\ell_\gamma}(\tau)$ the \textbf{"single-orbit term"}. Unless stated otherwise, $\gamma\in\mathcal{P}$ is fixed, and we write $\ell$ instead of $\ell_\gamma$ for convenience. Our main results are summarized as follows:
\begin{itemize}
\item \textbf{The $1-$Gevrey asymptotic expansion of the single-orbit term} (c.f. Definition \ref{def:borel}, Theorem \ref{thm:local}(i)).
We compute the uniform $1-$Gevrey asymptotic expansion of $U_{\ell}$ as $|\tau|\to\infty \text{ in } \mathbb H_+$:
\[U_\ell(\tau)\sim_1\widetilde{U}_\ell(\tau):=\sum_{j\geq0}a_j(\ell)\tau^{-j-\frac{1}{2}},
\text{ with }a_j(\ell):=\frac{(-1)^j}{2\sqrt{\pi}4^jj!}\sum_{k\geq1}b_k(\ell)(k\ell)^{2j}.
\]
Here $\widetilde U_\ell$ is a divergent formal series obtained by Taylor expansion of the exponential in $U_\ell$ in powers of $\tau^{-1}$ and formally interchanging the sums over the Taylor index $j$ and the iteration index $k$. We call it the \textbf{single-orbit series}. The asymptotic relation is justified by
\[
\left|U_\ell(\tau)-\sum_{j=0}^{N-1}a_j(\ell)\tau^{-j-\frac{1}{2}}\right|
\leq C_\ell^{\mathrm G}4^N N!|\tau|^{-N-\frac{1}{2}},\text{ for all }N\geq1 \text{ and }\tau\in\mathbb H_+.
\]
where $C_\ell^{\mathrm G}>0$ depends only on $\ell$.

\item \textbf{The Borel transform of the single-orbit series.}
In Theorem \ref{thm:local}(ii), we take the Borel transform (Definition \ref{def:borel}) over the single-orbit series, then we have $\BG\Utilde_\ell(\xi):=\sum_{j\geq0}\frac{a_j(\ell)}{\Gamma(j+\frac{1}{2})}\xi^{j-\frac{1}{2}}$, and denote it by $\Uhat_\ell(\xi)$. It converges in $0<|\sqrt{\xi}|<\frac{1}{2}$ and can be represented by
\[
 \Uhat_\ell(\xi)
 =\frac{1}{2\pi\sqrt\xi}\left(\sum_{k\geq1}b_k(\ell)-\xi\sum_{\substack{m\geq0\\ n\in\mathbb Z}}\frac{1}{\omega_{m,n}(\omega_{m,n}^2+\xi)}\right),
 \quad \omega_{m,n}:=m+\frac12+\frac{2\pi \ii n}{\ell}.
\]
This expression suggests that $\sqrt{\xi}\Uhat_\ell(\xi)$ can be meromorphically extended to $\mathbb{C}$ with pole set $\Omega_\ell=\{-\omega_{m,n}^2:m\geq0,\ n\in\mathbb{Z}\}$. In \cref{pro:cylinder}, the nonzero Borel singularities of the single-orbit series are identified with the resonance spectra of a hyperbolic cylinder, and their residues are related to the resonance spectrum multiplicities.

Thus, the questions can be answered: the spectral information emerges as the Borel singularities of the $1-$Gevrey asymptotic expansion of a single-orbit term. However, the singularities are related to the spectra of Laplace operator on the associated hyperbolic cylinders, rather than the spectra of $\Delta_X$.

\item \textbf{Stokes phenomenon for single-orbit series.}
 The original single-orbit term can be recovered by the Borel-Laplace transform of the single-orbit series along positive real-axis direction (\cref{thm:local}(ii)), where we have:
\[
 U_\ell(\tau)=\left(\mathcal{L}_{0}\circ \BG\Utilde_\ell\right)(\tau)=\int_0^\infty e^{-\tau\xi}\widehat{U}_\ell(\xi)\dd \xi,\quad \Re\tau>0.
\]
 More generally, we analyze all well-defined directions for Laplace transform of $\Uhat_\ell$. As Lemma \ref{lemma:singular direction} and Lemma \ref{lem:growth} show, in spite of infinite many singular directions in the Borel plane, the allowed directions for $\mathcal{L}_\theta \Uhat_\ell$ are almost everywhere. More precisely, the singular directions of $\Uhat_\ell$ in a fixed branch correspond to $\mathbb{Q}_{\mathrm{odd}}$, and all non-zero singular points in a fixed singular ray are $-(2j+1)^2\omega_{m_0,n_0}^2$, where $-\omega_{m_0,n_0}^2$ is the nearest singular point to the origin in the chosen singular ray. For almost every  non-singular ray and non-singular arc, the growth of $|\Uhat_\ell(\xi)|$ is of polynomial order as $|\xi|\to\infty$, which makes the Laplace integral well-defined on these regions.

 Thus, we can compute the limits of the differences of Laplace integrals on different directions approaching a singular ray, which implies the Stokes structure of the single-orbit series. As a result, it is represented by the summation of exponential terms about hyperbolic resonance spectra (\cref{thm:stokes}).
 \end{itemize}

\subsection{Organization of the Paper}

In Section~\ref{sec:local} we introduce the definitions and notation needed in resurgence theory firstly. Then we prove the $1-$Gevrey asymptotic relation between the single-orbit series and single-orbit term in the right half-plane. In the end, we compute the Borel transform of single-orbit series and describe its singular points in the two-sheeted Riemann surface. Section~\ref{section: cylinder} relates the singular points of the Borel transform of single-orbit series to the resonance spectra of the associated hyperbolic cylinders, which gives the spectrum interpretation of the singularities. Section~\ref{sec:stokes} is devoted to the analysis of the singular points of $\hat{U}_\ell(\xi)$ in different directions and the growth order of $\hat{U}_\ell(\xi)$ as $|\xi|\to\infty$, which allows us to study the Stokes phenomenon of single-orbit series in this section. Section~\ref{sec:conclusions} concludes the paper with several perspectives.

\section{Borel-Laplace Analysis of the Single-Orbit term}\label{sec:local}
In this section, we first introduce the notation and definitions in the resurgence theory\cite{ecallebook1,ecallebook2,ecallebook3}, then study the asymptotic expansion of the single-orbit term and the singularities of the Borel transform of the single-orbit series.

\subsection{$1-$Gevrey Expansion and Borel-Laplace Transform}
\begin{definition}[{\cite[Definitions 5.8, 5.21]{SauzinIntro}}]\label{def:borel}

We call a formal series $\widetilde{f}=\sum_{n\geq0}a_n\tau^{-n-\frac{1}{2}}$, where $a_n\in\mathbb{C}$, is a \textbf{$1-$Gevrey formal series}, if there exist $C,A>0$, such that $|a_n|\leq CA^n n!$.

Given an unbounded domain $\mathcal{D}\subset\mathbb{C}$ and a function $f:\mathcal{D}\to\mathbb{C}$, we say that the formal series $\widetilde{f}$ is the \textbf{uniform $1-$Gevrey asymptotic expansion} of $f$ in $\mathcal{D}$ and write $f(\tau)\sim_1\widetilde{f}(\tau)$ uniformly for $\tau\in \mathcal{D}$, if there exist $L,M>0$, such that for all $N\geq1$ and $\tau\in \mathcal{D}$,
\[\left|f(\tau)-\sum_{n=0}^{N-1}a_n\tau^{-n-\frac{1}{2}}\right|\leq LM^NN!|\tau|^{-N-\frac{1}{2}}.\]

The \textbf{Borel transform} and the \textbf{Laplace transform} are given by
\[\BG(\tau^{-\alpha})(\xi)=\frac{\xi^{\alpha-1}}{\Gamma(\alpha)},\quad \alpha>0,\quad \mathcal{L}_\theta\widehat{f}(\tau)=\int_0^{e^{\ii\theta}\infty}e^{-\tau\xi}\widehat{f}(\xi)\dd\xi,\]
whenever the integral converges. We apply $\BG$ to the half-integer series formally, then we get
\[\BG\left(\sum_{n\geq 0}a_n\tau^{-(n+\frac{1}{2})}\right)(\xi)=\sum_{n\geq 0}\frac{a_n}{\Gamma(n+\frac{1}{2})}\xi^{n-\frac{1}{2}}.\]
\end{definition}

\subsection{The $1-$Gevrey Expansion of the Single-Orbit term and Its Borel transform}

For any fixed $\ell>0$, the single-orbit term $U_\ell(\tau):=\frac{1}{\sqrt{4\pi\tau}}\sum_{k\geq1}b_k(\ell)e^{-\frac{(k\ell)^2}{4\tau}}$ is holomorphic on $\Hplus$, which is from the fact that
\[b_k(\ell)=\frac{\ell e^{-\frac{k\ell}{2}}}{1-e^{-k\ell}}\leq\frac{\ell}{1-e^{-\ell}}e^{-\frac{k\ell}{2}},\text{ and } \Re\left(\frac{1}{\tau}\right)=\Re\left(\frac{\bar{\tau}}{|\tau|^2}\right)>0.\]
We expand the exponential term in $U_\ell(\tau)$ in powers of $\tau^{-1}$, then
\[U_\ell(\tau)=\frac{1}{2\sqrt{\pi}\tau^{\frac{1}{2}}}   \sum_{k\geq1}b_k(\ell) \sum_{j\geq0}\frac{(-1)^j}{j!}\left(\frac{(k\ell)^2}{4\tau}\right)^j.\]
Subsequently, by formally interchanging the summation over index $j$ and $k$ in $U_\ell(\tau)$, we will get a formal series and call it the "single-orbit series", it is
\[
\widetilde{U}_\ell(\tau):=\sum_{j\geq0}a_j(\ell)\tau^{-j-\frac{1}{2}},\text{ where }
a_j(\ell):=\frac{(-1)^j}{2\sqrt{\pi}4^j j!}\sum_{k\geq1}b_k(\ell)(k\ell)^{2j}.
\]
Since $\sum_{k\geq1}b_k(\ell)(k\ell)^{2j}<\infty$ for any fixed $j$, each coefficient $a_j(\ell)$ is well defined. However, formal index interchange does not guarantee the equality of the two series. Even worse, different from the holomorphic  $U_\ell(\tau)$ in the half-plane $\Hplus$, the series $\widetilde{U}_\ell(\tau)$ is divergent, as the following Lemma shows.

\begin{lemma}{\label{lemma:single series divegent}}
For each fixed $\ell>0$, the formal series $\widetilde{U}_\ell$ diverges for any $\tau\in\Hplus$.
\end{lemma}
\begin{proof}
For $j\geq 1$, choosing the term $k_j=\lceil \frac{4j}{\ell}\rceil$ in the summation and using $4j\leq k_j\ell<4j+\ell$, we obtain that
\[
\begin{aligned}
|a_j(\ell)|
&=\frac{1}{2\sqrt{\pi}4^j j!}
  \sum_{k\geq1}b_k(\ell)(k\ell)^{2j}\\
&\geq \frac{\ell}{2\sqrt{\pi}4^j j!}
  \sum_{k\geq1} e^{-\frac{k\ell}{2}}(k\ell)^{2j}\\
&\geq \frac{\ell}{2\sqrt{\pi}4^j j!} e^{-\frac{k_j\ell}{2}}(k_j\ell)^{2j}\\
&\geq \frac{\ell e^{-\frac{\ell}{2}}}{2\sqrt{\pi}}4^j j!\frac{j^{2j}e^{-2j}}{(j!)^2}.
\end{aligned}
\]
By Stirling's formula, $\frac{j^{2j}e^{-2j}}{(j!)^2}\sim\frac{1}{2\pi j}$ as $j\to\infty$. Thus, there exists $J>0$, such that for any $j>J$,
\[
|a_j(\ell)|\geq c_\ell\frac{4^j j!}{j},
\]
where $c_\ell:=\frac{\ell e^{-\frac{\ell}{2}}}{8\pi\sqrt{\pi}}$. Thus, for any fixed $\tau\neq0$, $|a_j(\ell)\tau^{-j-\frac{1}{2}}|\geq c_\ell \frac{4^j j!}{j|\tau|^{j+\frac{1}{2}}}$, which tends to infinity as $j\to\infty$, so $\widetilde{U}_\ell$ diverges for any $\tau\in\Hplus$.
\end{proof}

Although the single-orbit series $\widetilde{U}_\ell$ is divergent, it is possible to describe the large-time asymptotic behavior of the single-orbit term $U_\ell$ by $\widetilde{U}_\ell$. The following theorem claims that $\widetilde{U}_\ell$ is the $1-$Gevrey asymptotic expansion of $U_\ell$ as $|\tau|\to\infty$ in $\Hplus$. Moreover, we study the singularities of its Borel transform. The locations of these singularities are related to the resonance spectra of the associated hyperbolic cylinder (see \cref{pro:cylinder}).

\begin{theorem}\label[theorem]{thm:local}
Let $\ell>0$ and $U_\ell$ as defined in \eqref{eq:objects}. Then we have:
\begin{itemize}
\item[(i)] As $|\tau|\to\infty$ in $\Hplus$, the holomorphic function $U_\ell$ has the uniform $1-$Gevrey asymptotic expansion
\[
 U_\ell\sim_1\widetilde{U}_\ell:=\sum_{j\geq0}a_j(\ell)\tau^{-j-\frac{1}{2}},
 \quad a_j(\ell)=\frac{(-1)^j}{2\sqrt{\pi}4^j j!}\sum_{k\geq1}b_k(\ell)(k\ell)^{2j}.
\]
 Furthermore, $\widetilde{U}_\ell$ is a $1-$Gevrey formal series. More precisely, for any $N\geq1$ and $\tau\in\Hplus$,
\[
 \left|U_\ell(\tau)-\sum_{j=0}^{N-1}a_j(\ell)\tau^{-j-\frac{1}{2}}\right|
 \leq |a_N(\ell)|\,|\tau|^{-N-\frac{1}{2}}
 \leq \frac{e^{\frac{\ell}{2}}}{\sqrt{\pi}(1-e^{-\ell})}4^N N!|\tau|^{-N-\frac{1}{2}}.
\]
\item[(ii)] The Borel transform of $\widetilde{U}_\ell$ can be represented as
\begin{equation}\label{eq:local-ML}
\widehat{U}_\ell(\xi)=\frac{1}{2\pi\sqrt{\xi}}\left(c_0(\ell)-\xi\sum_{m\geq0,n\in\mathbb{Z}}\frac{1}{\omega_{m,n}(\omega_{m,n}^2+\xi)}\right),\quad 0<|\sqrt{\xi}|<\frac{1}{2}.
\end{equation}
where $c_0(\ell):=\sum_{k\geq1}b_k(\ell)$ and $\omega_{m,n}:=m+\frac{1}{2}+\frac{2\pi\ii n}{\ell}$ for all $m\in\mathbb{Z}_{\geq 0},n\in\mathbb{Z}$. Furthermore, denote a new series $M_\ell(\xi):=\sqrt{\xi}\widehat{U}_\ell(\xi)$ in the same region, it can be extended meromorphically to $\mathbb{C}$, with poles at $\Omega_\ell:=\{-\omega_{m,n}^2:m\geq0,\ n\in\mathbb{Z}\}$ and
\[
\Res\left(M_\ell(\xi);\xi=-\omega_{m,n}^2\right)=\frac{\omega_{m,n}}{2\pi},\text{ for all }-\omega_{m,n}^2\in\Omega_\ell.
\]
Moreover, the single-orbit term can be recovered by taking the Borel-Laplace integration along the direction $\theta=0$ of the single-orbit series. In other words,
\begin{equation}\label{eq:local-borel}
U_\ell(\tau)= \mathcal{L}_0\BG \widetilde{U}_\ell(\tau)
=\int_0^\infty e^{-\tau\xi}\widehat{U}_\ell(\xi)\dd\xi,\quad \tau\in\Hplus.
\end{equation}

\end{itemize}
\end{theorem}

\begin{proof}
\begin{itemize}
\item[(i)] Consider the Taylor expansion of the exponential term in $U_\ell(\tau)$ and denote $r_{N,k}(\tau):=e^{-\frac{(k\ell)^2}{4\tau}}-\sum_{j=0}^{N-1}\frac{(-\frac{(k\ell)^2}{4\tau})^j}{j!}$, we have
\[
U_\ell(\tau)=\frac{1}{2\sqrt{\pi\tau}}\sum_{k\geq1}b_k(\ell)\left(\sum_{j=0}^{N-1}\frac{(-1)^j}{j!}\left(\frac{(k\ell)^2}{4\tau}\right)^j+r_{N,k}(\tau)\right).
\]
Notice that for each fixed $N$, $\sum_{k\geq1}b_k(\ell)(k\ell)^{2N}$ is finite, thus the finite sum and the infinite sum can be interchanged. It gives
\[
U_\ell(\tau)
=\frac{1}{2\sqrt{\pi\tau}}
\sum_{j\geq0}^{N-1}\sum_{k\geq1}b_k(\ell)\frac{(-1)^j}{j!}\left(\frac{(k\ell)^2}{4\tau}\right)^j
+\frac{1}{2\sqrt{\pi\tau}}\sum_{k\ge1}b_k(\ell)r_N(\tau).
\]
Denote the remainder with $R_N(\tau):=\frac{1}{2\sqrt{\pi\tau}}\sum_{k\geq 1}b_k(\ell)r_{N,k}(\tau)$, and observe that for $\Re\tau>0$, $|r_{N,k}(\tau)|\leq\frac{|\frac{(k\ell)^2}{4\tau}|^N}{N!}$. Then,
\[
 U_\ell(\tau)=\sum_{j=0}^{N-1}a_j(\ell)\tau^{-j-\frac{1}{2}}+R_N(\tau),
 \quad
 a_j(\ell)=\frac{(-1)^j}{2\sqrt{\pi}4^j j!}\sum_{k\geq 1}b_k(\ell)(k\ell)^{2j},
\]
where
\[
|R_N(\tau)|
\leq \frac{|\tau|^{-N-\frac{1}{2}}}{2\sqrt{\pi}4^N N!}\sum_{k\geq 1}b_k(\ell)(k\ell)^{2N}
=|a_N(\ell)|\,|\tau|^{-N-\frac{1}{2}},\quad \tau\in\Hplus.
\]

Now, to prove the $1-$Gevrey property, we estimate $|a_N(\ell)|$.
For any $x\in[k\ell,(k+1)\ell]$, there is $(k\ell)^{2N}e^{-k\ell/2}\leq e^{\ell/2}x^{2N}e^{-x/2}$. Integrating over these intervals and summing in $k$, we obtain that
\[
 \sum_{k\geq1}b_k(\ell)(k\ell)^{2N}
 \leq \frac{\ell}{1-e^{-\ell}}\sum_{k\ge1}e^{-\frac{k\ell}{2}}(k\ell)^{2N}
 \leq\frac{e^{\frac{\ell}{2}}}{1-e^{-\ell}}\int_0^\infty x^{2N}e^{-\frac{x}{2}}\dd x
 =\frac{e^{\frac{\ell}{2}}}{1-e^{-\ell}}2^{2N+1}(2N)!.
\]
Thus, by $\binom{2N}{N}\leq4^N$,
\[
 |a_N(\ell)|
 \leq\frac{e^{\frac{\ell}{2}}}{\sqrt{\pi}(1-e^{-\ell})}\frac{(2N)!}{N!}
 \leq \frac{e^{\frac{\ell}{2}}}{\sqrt{\pi}(1-e^{-\ell})}4^N N!.
\]
Hence, the series $\widetilde{U}_\ell$ is a $1-$Gevrey formal series and it is the uniform $1-$Gevrey asymptotic expansion of $U_\ell$ as $|\tau|\to\infty$ in $\Hplus$.

\item[(ii)] By the definition of the Borel transform and the identity
$j!\Gamma(j+\frac{1}{2})=\sqrt{\pi}\frac{(2j)!}{4^j}$, we have
\[
\widehat{U}_\ell(\xi)
=\sum_{j\geq0}\frac{a_j(\ell)}{\Gamma(j+\frac{1}{2})}\xi^{j-\frac{1}{2}}
=\frac{1}{2\pi\sqrt{\xi}}\sum_{j\geq0}\sum_{k\geq1}b_k(\ell)\frac{(-1)^j(k\ell)^{2j}\xi^j}{(2j)!}.
\]
Recall that $b_k(\ell)\leq\frac{\ell e^{-\frac{k\ell}{2}}}{1-e^{-\ell}}$, then
\begin{equation}\label{eqn: estimate for Uhat}
\sum_{k\geq1}b_k(\ell)\sum_{j\geq0}\frac{(k\ell)^{2j}|\xi|^j}{(2j)!}
<\sum_{k\geq1}b_k(\ell)e^{k\ell\sqrt{|\xi|}}
\leq \frac{\ell}{1-e^{-\ell}}\sum_{k\geq1}e^{-k\ell(\frac{1}{2}-\sqrt{|\xi|})}
<\infty
\end{equation}
for $0<|\xi|<1/4$. Thus, the series for $\widehat{U}_\ell$ converges absolutely in $\{\xi\in\mathbb{C}:0<|\xi|<1/4\}$ and the double-summation over index $j,k$ can be interchanged in the same domain. Interchanging the index $j,k$ in $\Uhat_\ell$ and summing the Taylor series of the function $\cos$, there is
\begin{equation}{\label{eqn:borel transform of Uell with cos}}
 \widehat{U}_\ell(\xi)
 =\frac{1}{2\pi\sqrt\xi}\sum_{k\ge1}b_k(\ell)\cos(k\ell\sqrt\xi),
 \quad 0<|\xi|<\frac{1}{4}.
\end{equation}
Notice that $|\cos(k\ell\sqrt{\xi})|\leq e^{k\ell|\Im\sqrt{\xi}|}$, retaining the similar estimate in (\ref{eqn: estimate for Uhat}), the range in equation (\ref{eqn:borel transform of Uell with cos}) can be extended to $\{\xi\in\mathbb{C}:|\Im \sqrt{\xi}|<\frac{1}{2}\}\setminus\{0\}$ while preserving the identity. 

We now prove \eqref{eq:local-ML}. Since $b_k(\ell)=\ell\sum_{m\geq0}e^{-(m+\frac{1}{2})k\ell}$, we have
\[
\begin{aligned}
2\pi M_\ell(\xi)-c_0(\ell)
&=\sum_{k\geq1}b_k(\ell)
\left(\cos(k\ell\sqrt{\xi})-1\right)\\
&=\ell\sum_{m\geq0}\sum_{k\geq1}e^{-(m+\frac{1}{2})k\ell}\left(\cos(k\ell\sqrt{\xi})-1\right).
\end{aligned}
\]
To compute the summation over index $k$, we use the identity
\[
\frac{\ell}{2}\coth\left(\frac{\ell w}{2}\right)
=\frac{\ell}{2}+\ell\sum_{k\geq1}e^{-k\ell w},
\quad \Re w>0,
\]
and the classical expansion (\cite[Formula 1.421(4)]{GradshteynRyzhik})
\[
\frac{\ell}{2}\coth\left(\frac{\ell w}{2}\right)
=\lim_{N\to\infty}\sum_{n=-N}^{N}\frac{1}{w+2\pi \ii n/\ell},
\]
where the latter sum is understood in the symmetric sense.

Applying these identities with $w=m+\frac{1}{2}\pm \ii\sqrt{\xi}$ and $w=m+\frac{1}{2}$, we obtain that
\[
\begin{aligned}
&\ell\sum_{k\geq1}e^{-(m+\frac{1}{2})k\ell}\left(\cos(k\ell\sqrt{\xi})-1\right)\\
&=\frac{\ell}{2}\sum_{k\geq1}e^{-(m+\frac{1}{2})k\ell}\left(e^{\ii k\ell\sqrt{\xi}}+e^{-\ii k\ell\sqrt{\xi}}-2\right)\\
&=\frac{\ell}{4}\left(\coth\left(\frac{\ell}{2}(m+\frac{1}{2}-\ii\sqrt{\xi})\right)
+\coth\left(\frac{\ell}{2}(m+\frac{1}{2}+\ii\sqrt{\xi})\right)
-2\coth\left(\frac{\ell}{2}(m+\frac{1}{2})\right)\right)\\
&=\frac{1}{2}\sum_{n\in\mathbb{Z}}\left(\frac{1}{\omega_{m,n}-\ii\sqrt{\xi}}+\frac{1}{\omega_{m,n}+\ii\sqrt{\xi}}-\frac{2}{\omega_{m,n}}\right)\\
&=-\xi\sum_{n\in\mathbb{Z}}\frac{1}{\omega_{m,n}(\omega_{m,n}^2+\xi)}.
\end{aligned}
\]
Thus,
\[
2\pi M_\ell(\xi)-c_0(\ell)
=-\xi\sum_{\substack{m\geq0\\n\in\mathbb{Z}}}\frac{1}{\omega_{m,n}(\omega_{m,n}^2+\xi)}.
\]
Since $M_\ell(\xi)=\sqrt{\xi}\widehat{U}_\ell(\xi)$, this proves \eqref{eq:local-ML} for $0<|\sqrt{\xi}|<\frac{1}{2}$.

We next prove the meromorphic continuation.
For $\xi$ in a fixed compact subset of $\mathbb {C}\setminus\Omega_\ell$ and any sufficiently large $|\omega_{m,n}|$, there is
\[
\left|\frac{\xi}{\omega_{m,n}(\omega_{m,n}^2+\xi)}\right|
\leq C|\omega_{m,n}|^{-3}.
\]
Furthermore,
$$
\begin{aligned}
\sum_{n\in\mathbb{Z}}|\omega_{m,n}|^{-3}
&=\sum_{n\in\mathbb{Z}}\left((m+\frac{1}{2})^2+(\frac{2\pi n}{\ell})^2\right)^{-3/2}\\
&\leq \frac{1}{(m+\frac{1}{2})^3}+2\int_0^{\infty} \left((m+\frac{1}{2})^2+(\frac{2\pi x}{\ell})^2\right)^{-3/2}\dd x\\
&=\frac{1}{(m+\frac{1}{2})^3}+\frac{\ell}{\pi(m+\frac{1}{2})^2}.
\end{aligned}
$$
Thus,
\[
\sum_{m\geq0}\sum_{n\in\mathbb{Z}}|\omega_{m,n}|^{-3}<\infty.
\]
Then the series in \eqref{eq:local-ML} converges uniformly on the compact subsets of $\mathbb C\setminus\Omega_\ell$, and $M_\ell(\xi)$ can be extended meromorphically to the whole complex plane.

Since $\Re\omega_{m,n}>0$, distinct pairs $(m,n)$ give different values of $-\omega_{m,n}^2$. Hence all of these poles are simple, and
\[
\Res\left(M_\ell(\xi); \xi=-\omega_{m,n}^2\right)
=-\frac{1}{2\pi}\Res\left(\frac{\xi}{\omega_{m,n}(\omega_{m,n}^2+\xi)}; \xi=-\omega_{m,n}^2\right)
=\frac{\omega_{m,n}}{2\pi}.
\]
Finally, we establish the Borel-Laplace relation between $U_\ell$ and $\widetilde{U}_\ell$. By equation (\ref{eqn:borel transform of Uell with cos}), for any $\xi\in\mathbb{R}^+$, $|\widehat{U}_\ell(\xi)|\leq\frac{1}{2\pi\sqrt{\xi}}\sum_{k\geq1}b_k(\ell)
=\frac{c_0(\ell)}{2\pi\sqrt{\xi}}$. Thus the Laplace integral converges absolutely when $\Re\tau>0$, and we can interchange the summation with the integral. Then
\[
\begin{aligned}
\int_0^\infty e^{-\tau\xi} \widehat{U}_\ell(\xi)\dd\xi
&=\frac{1}{2\pi}\sum_{k\geq1}b_k(\ell) \int_0^\infty e^{-\tau\xi}\xi^{-\frac{1}{2}}\cos(k\ell\sqrt{\xi})\dd\xi\\
&=\frac{1}{\sqrt{4\pi\tau}}\sum_{k\geq1}b_k(\ell)e^{-\frac{(k\ell)^2}{4\tau}}\\
&=U_\ell(\tau),
\end{aligned}
\]
which proves \eqref{eq:local-borel}.
\end{itemize}
\end{proof}

\begin{remark}\label{Remark:geomety distributionsingle}
Denote $\xi=x+\ii y$, we observe that the singular points of $\widehat{U}_\ell(\xi)$  are the intersections of two families of parabolas:
\[
\begin{aligned}
\mathcal{P}_m^{R}:\quad
&y^2=4\left(m+\frac{1}{2}\right)^2\left(x+\left(m+\frac{1}{2}\right)^2\right),\quad m\geq0;\\
\mathcal{P}_n^{L}:\quad
&y^2=4\left(\frac{2\pi n}{\ell}\right)^2 \left(\left(\frac{2\pi n}{\ell}\right)^2-x\right),\quad n\geq0.
\end{aligned}
\]
 For $n=0$, the second equation degenerates to $y=0$, with the poles $-(m+\frac{1}{2})^2$ on the negative real axis. Thus, $\widehat{U}_\ell(\xi)$ is meromorphic on the two-sheeted Riemann surface of $\sqrt{\xi}$ with non-zero poles at the intersections of $\{\mathcal{P}_m^{R}\}_{m\in\mathbb{Z}_{>0}}$ and $\{\mathcal{P}_n^{L}\}_{n\in\mathbb{Z}_{\geq 0}}$, as shown in Figure \ref{fig:borel-singularities}.

\begin{figure}[htbp]
\centering

\begin{minipage}[t]{0.48\textwidth}
    \centering
    \vspace{0pt}
    \parbox[c][5.5cm][c]{\linewidth}{
        \centering
        \includegraphics[
            width=0.82\linewidth,
            height=5.3cm,
            keepaspectratio
        ]{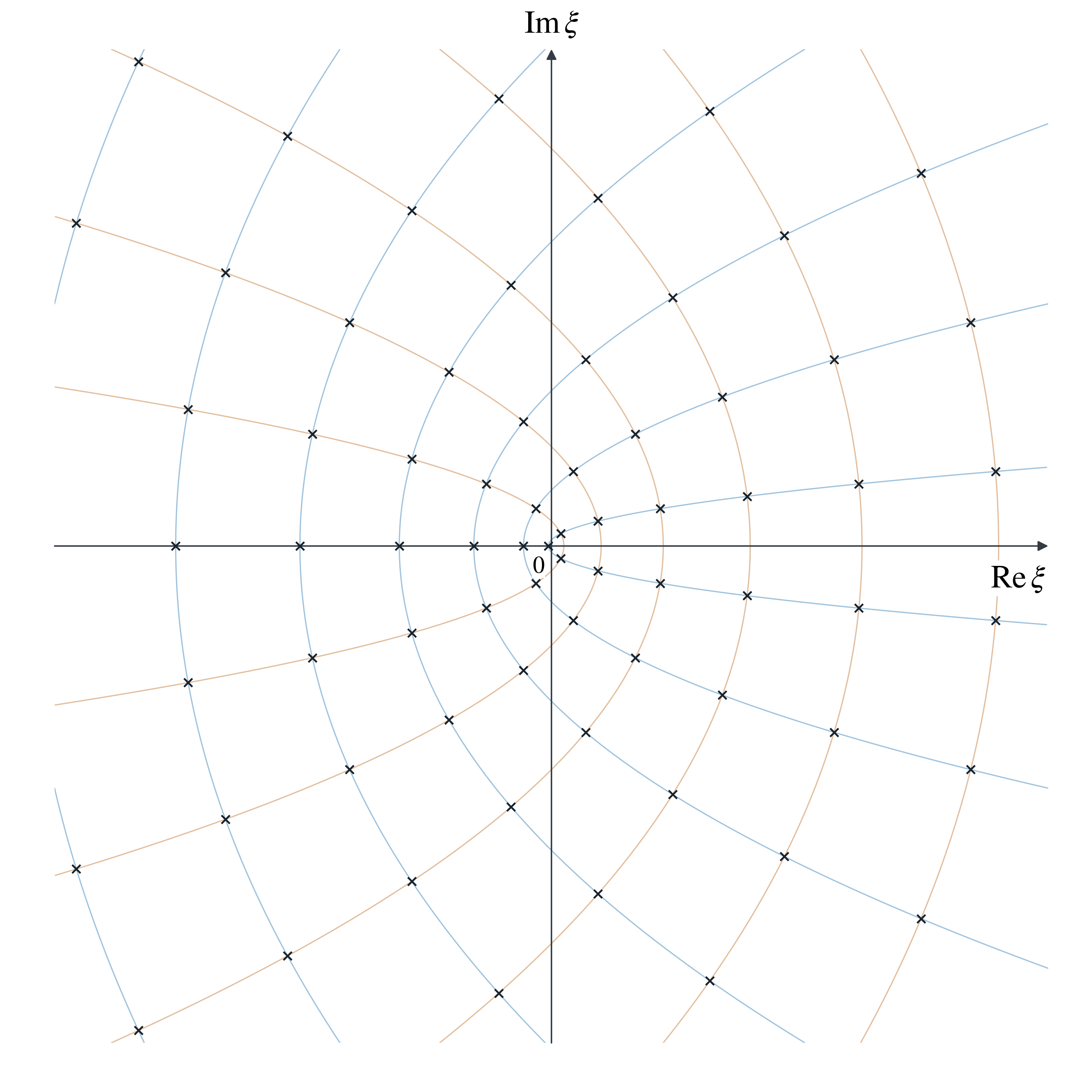}
    }

    \smallskip
    (a)
\end{minipage}
\hfill
\begin{minipage}[t]{0.48\textwidth}
    \centering
    \vspace{0pt}
    \parbox[c][5.5cm][c]{\linewidth}{
        \centering
        \includegraphics[
            width=\linewidth,
            height=5.3cm,
            keepaspectratio
        ]{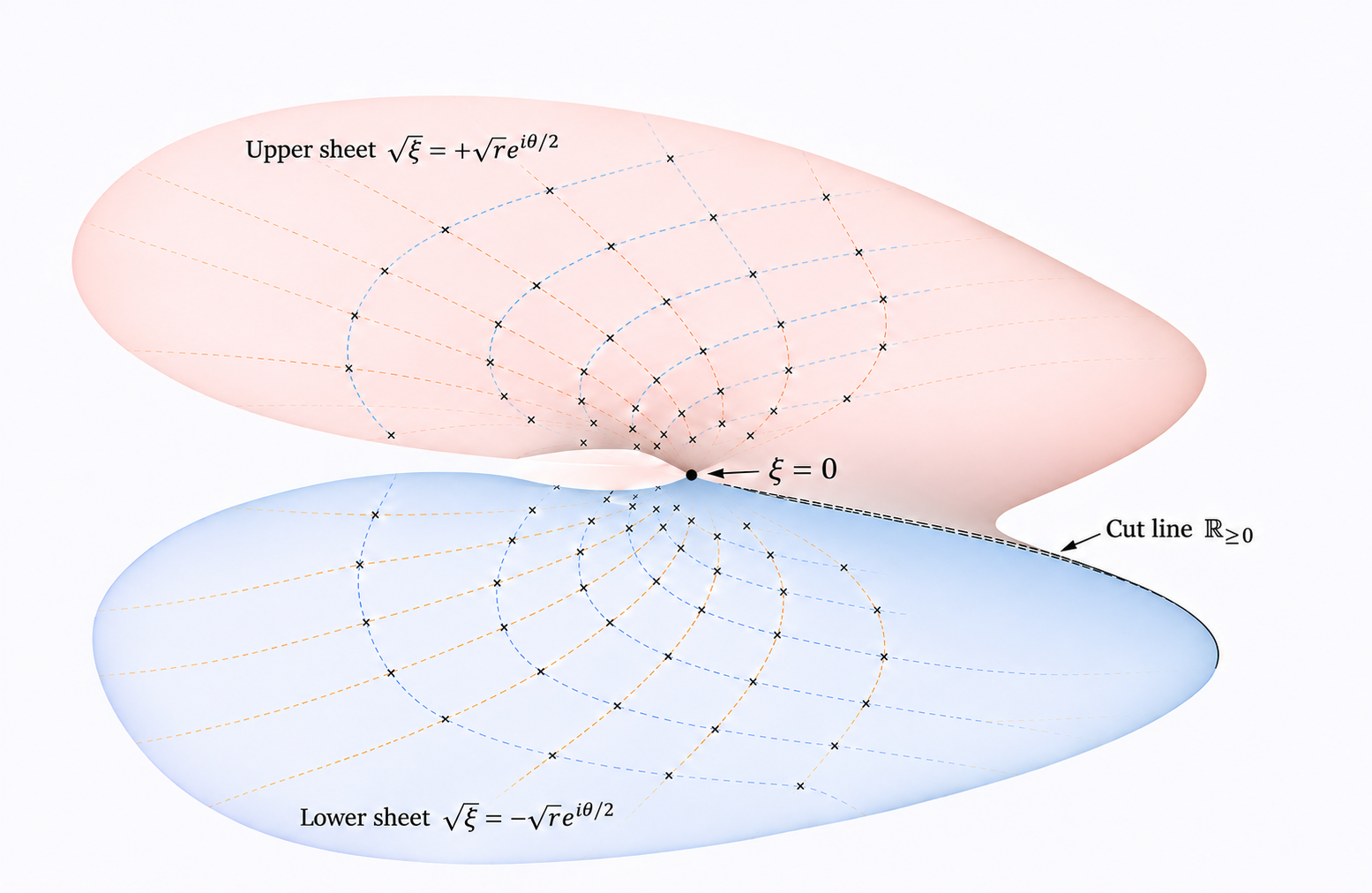}
    }

    \smallskip
    (b)
\end{minipage}

\caption{
Geometric distribution of Borel singularities.
(a) The singular points $\xi=-\omega_{m,n}^2$ on a fixed sheet are the intersections of two families of parabolas in the Borel plane.
(b) The two-sheeted Riemann surface of $\widehat{U}_{\ell}$, showing the singularities on both sheets and the branch point at $\xi=0$.
}
\label{fig:borel-singularities}
\end{figure}

\end{remark}

\section{Cylinder resonance spectrum and Borel Singularity}\label{section: cylinder}

In this section, we identify the nonzero Borel singularities of the single-orbit series with the resonance spectra of a hyperbolic cylinder and relate their residues to the resonance multiplicities.

Recall the classical results of the resonance spectrum set of a hyperbolic cylinder (\cite[Theorem 1.2]{APW} and \cite[Section 2,3]{BJP}). Let $C_\ell$ be the hyperbolic cylinder obtained from $\mathbb{H}$ by identifying $z$ with $e^\ell z$. Its resonance spectra are defined as the poles of the meromorphic continuation of $(\Delta_{C_\ell}-s(1-s))^{-1}$, in another form, they also be the zeros of the Selberg-zeta function in the hyperbolic cylinder (\cite[Section 2,3]{BJP}). This function extends to an entire function, and for $\Re s>1$, is given by
\[Z_{C_\ell}(s)=\prod_{m\geq0}(1-e^{-(s+m)\ell})^2.\]

Thus, the resonance spectrum set of the hyperbolic cylinder is
\begin{equation}\label{eq:cylinder}
 \mathcal{R}(C_\ell)=-\mathbb{Z}_{\geq 0}+\frac{2\pi\ii}{\ell}\mathbb{Z},\quad\text{ with multiplicities } m_{C_\ell}(s)=2,\text{ for all }s\in\mathcal{R}(C_\ell).
\end{equation}

Accounting for the two orientations of the geodesic orbit, we have $\mathcal{P}_{C_\ell}=\{\gamma,\gamma^{-1}\}$. Thus the sum of two primitive-orbit contributions on the cylinder is
$$U_{\ell_\gamma}+U_{\ell_{\gamma^{-1}}}=2U_{\ell_\gamma},$$
we denote this sum by $2U_\ell$. The following proposition identifies the cylinder resonance spectra with the non-zero Borel singularities of the single-orbit series $2U_\ell$.  

\begin{proposition}\label[proposition]{pro:cylinder}
Let $\ell>0$ be the length of a primitive closed
geodesic in $X$. The map
\[
\begin{aligned}
A:\mathcal R(C_\ell)&\longrightarrow\Omega_\ell,\\
s&\longmapsto -(s-\frac{1}{2})^2
\end{aligned}
\]
is a bijection between the resonance spectrum set of
$C_\ell$ and the set of nonzero poles of the Borel transform of the single-orbit series $2\Utilde_\ell$. Fix $\sigma\in\{1,-1\}$ and choose the branch $\sqrt{re^{\ii\theta}}=\sigma r^{\frac{1}{2}}e^{\ii\theta/2}$, then, for any $s\in\mathcal{R}(C_\ell)$,
\[2\pi\ii\Res\left(2\widehat{U}_\ell(\xi); \xi=A(s)\right)=\sigma m_{C_\ell}(s).\]
\end{proposition}

\begin{proof}
By \eqref{eq:cylinder}, all points in $\mathcal{R}(C_\ell)$ can be written as $s_{m,n}=\frac{1}{2}-\omega_{m,n}$, and $A(s_{m,n})=-\left(s_{m,n}-\frac{1}{2}\right)^2=-\omega_{m,n}^2$. Hence $A$ is surjective. If $A(s)=A(t)$, then $s=t$ or $s+t=1$. The latter is impossible because $\Re s,\Re t\leq0$, then $A$ is injective.

Since $\Re\omega_{m,n}>0$, the chosen branch gives $\sqrt{-\omega_{m,n}^2}=\sigma \ii\omega_{m,n}$. Using the residue formula for $M_\ell(\xi)=\sqrt{\xi}\widehat{U}_\ell(\xi)$, we have
\[
\Res\left(2\widehat{U}_\ell(\xi); \xi=-\omega_{m,n}^2\right)
=\frac{1}{\sigma \ii\omega_{m,n}}\Res\left(2M_\ell(\xi);\xi=-\omega_{m,n}^2\right)
=\frac{2\omega_{m,n}}{2\pi\sigma \ii\omega_{m,n}}
=\frac{2\sigma}{2\pi \ii}.
\]
Together with $m_{C_\ell}(s)=2$, the residue identity is proved.
\end{proof}

\begin{remark}
The poles and residues of the meromorphic spectral density of a hyperbolic cylinder were computed by Marklof\cite[Proposition 8.2]{Marklof}. Proposition \ref{pro:cylinder} identifies this classical resonance structure with the nonzero Borel singularities of the formal single-orbit series. It provides the resonance spectrum a new interpretation in resurgence theory. Beyond this correspondence, the Borel transform allows us to study the Laplace integration along different directions and the associated Stokes phenomenon. In particular, the cylinder resonance spectra determine the Stokes phenomenon as shown in Section \ref{sec:stokes}.
\end{remark}

\section{The Stokes Phenomenon of the single-orbit Series}\label{sec:stokes}
In this section, we study the Stokes phenomenon of the single-orbit series, which is described  by the differences between Laplace integrals of the Borel transform $\widehat U_\ell$ along different directions. Although the Borel transform extends meromorphically, its singular directions are dense, as shown in Lemma \ref{lemma:singular direction}, so the convergence of the corresponding Laplace integrals is nontrivial. We establish growth estimates for $\widehat U_\ell$ in Lemma \ref{lem:growth} and use them to describe the Stokes phenomenon when crossing a singular ray (Theorem \ref{thm:stokes}).

\begin{definition}
We say a ray in the Borel-plane $\ell_\theta:=\{\xi\in\mathbb{C}:|\xi|>0,\arg\xi=\theta\}$ is a \textbf{singular direction}, if there exists some $-\omega_{m,n}^2\in\Omega_{\ell}$ located in $\ell_\theta$; on the contrary, a ray in the Borel plane is a \textbf{non-singular direction} if it avoids any singular points in $\Omega_\ell$.
\end{definition}

In the following lemma, we describe the distribution of the singular directions and the singular points on these directions.
\begin{lemma}{\label{lemma:singular direction}}
Consider the singular directions in $(0,2\pi)$.
\begin{itemize}
\item[(i)]
The singular directions $1-1$ correspond to $\mathbb{Q}_{\mathrm{odd}}$, where
$$\mathbb{Q}_{\mathrm{odd}}:=\left\{\frac{p}{q}\in\mathbb{Q}:\gcd(|p|,q)=1,q\text{ is a positive odd integer}\right\}.
$$
Consequently, the singular directions are countable and dense, whereas the non-singular directions are dense and have full Lebesgue measure in $(0,2\pi)$.

\item[(ii)]
Let $\theta\in(0,2\pi)$ be a singular direction, and let $-\omega_{m_0,n_0}^2$ be the pole of smallest modulus on the ray with $\arg\xi=\theta$. Then $2m_0+1$ and $|n_0|$ are coprime, and
\[\Omega_\ell\cap\{\arg\xi=\theta\}=\left\{-(2j+1)^2\omega_{m_0,n_0}^2:j\geq0\right\}.\]
\end{itemize}
\end{lemma}

\begin{proof}
(i) Since $\omega_{m,n}=m+\frac{1}{2}+\frac{2\pi \ii n}{\ell}$, we have $\arg(-\omega_{m,n}^2)=\pi+2\arctan\left(\frac{4\pi n}{\ell(2m+1)}\right)$. Hence $\theta\in(0,2\pi)$ is singular if and only if
\[\frac{\ell}{4\pi}\tan\left(\frac{\theta-\pi}{2}\right)=\frac{n}{2m+1},\]
which gives the correspondence between the set of singular directions and $\mathbb{Q}_{\mathrm{odd}}$.

(ii) Two poles lie on the same ray if and only if $n(2m_0+1)=n_0(2m+1)$. If $d=\gcd(2m_0+1,|n_0|)>1$, then $d$ is odd and $\omega_{m_0,n_0}=d\omega_{\frac{m_0+\frac{1}{2}}{d}-\frac{1}{2},\frac{n_0}{d}}$, which contradicts the minimality of $|\omega_{m_0,n_0}|$. Thus $d=1$.

From $n(2m_0+1)=n_0(2m+1)$ and the fact that $2m_0+1,|n_0|$ are coprime, we have
\[
2m+1=k(2m_0+1),\quad n=kn_0
\]
for some positive odd integer $k$.
\end{proof}

Since the singular directions are dense, no nonsingular ray admits a holomorphic sectorial neighbourhood. However, the local finiteness of the singularities ensures that each nonsingular ray has a holomorphic neighbourhood whose width may vary with the radius. Therefore, we estimate the growth of $\widehat U_\ell$ along almost every ray and on a sequence of large circles. These estimates ensure the absolute convergence of the corresponding Laplace integrals in suitable half-planes and provide the control needed for contour deformation in Theorem \ref{thm:stokes}.

\begin{lemma}\label{lem:growth}
\begin{itemize}
\item[(i)]
For almost every direction $\alpha\in(0,2\pi)$,
\[
\widehat{U}_\ell(re^{\ii\alpha})=O_{\ell,\alpha}(r^3),\quad r\to\infty.
\]
Then $\mathcal{L}_\alpha\widehat{U}_\ell(\tau)$ converges absolutely whenever $\Re(\tau e^{\ii\alpha})>0$.

\item[(ii)]
For any $N\in\mathbb{Z}_{\geq 1}$, there exists $R_N\in[N,N+1]$ such that
\[
\sup_{|\xi|=R_N}|\widehat{U}_\ell(\xi)|=O_\ell(R_N^2).
\]
\end{itemize}
\end{lemma}

\begin{proof}
Fix $R>1$. For $|\xi|=R$,  the series $M_\ell(\xi)=\frac{1}{2\pi}\left[c_0(\ell)-\xi\sum_{m,n}\frac{1}{\omega_{m,n}(\omega_{m,n}^2+\xi)}\right]$
can be split into three series, with the sum taken over $|\omega_{m,n}|^2\leq \frac{R}{2}$, $\frac{R}{2}<|\omega_{m,n}|^2<2R$ and $|\omega_{m,n}|^2\geq2R$ respectively. Since $|\omega_{m,n}|^2=\left(m+\frac{1}{2}\right)^2+\frac{4\pi^2n^2}{\ell^2}$,
we have
\[
\#\{(m,n):|\omega_{m,n}|\leq R\}=O_\ell(R^2) \text{ and }\#\{(m,n):R\leq|\omega_{m,n}|<2R\}=O_\ell(R^2).
\]
Decomposing the plane into different annulus with radius $2^jR$, there exist positive constants $c_\ell^1$ and $c_\ell^2$ such that
\[
\sum_{|\omega_{m,n}|\leq R}|\omega_{m,n}|^{-1}
=\sum_{j\ge0}\sum_{2^{-j-1} R<|\omega_{m,n}|<2^{-j}R}|\omega_{m,n}|^{-1}
\leq\sum_{j\ge0}
\frac{c_\ell^1(2^{-j} R)^2}{2^{-j-1}R}
=O_\ell(R),
\]
and
\[
\sum_{|\omega_{m,n}|\geq R}|\omega_{m,n}|^{-3}
=\sum_{j\geq 0}\sum_{2^j R<|\omega_{m,n}|<2^{j+1}R}|\omega_{m,n}|^{-3}
\leq \sum_{j\geq 0}\frac{c_\ell^2(2^{j+1} R)^{2}}{(2^j R)^{3}}
=O_\ell(R^{-1}).
\]

Thus, for $|\omega_{m,n}|^2\leq R/2$, $|\omega_{m,n}^2+\xi|\geq |\xi|-|\omega_{m,n}^2|\geq\frac{R}{2}$, we have 
\[
\sum_{|\omega_{m,n}|^2\leq R/2}\left|\frac{\xi}{\omega_{m,n}(\omega_{m,n}^2+\xi)}\right|
\leq 2\sum_{|\omega_{m,n}|\leq\sqrt{R/2}}|\omega_{m,n}|^{-1}
=O_\ell(\sqrt R).
\]
And when $|\omega_{m,n}^2|>2R$, $|\omega_{m,n}^2+\xi|\geq |\omega_{m,n}^2|-|\xi|\geq\frac{|\omega_{m,n}^2|}{2}$, which makes
\[
\sum_{|\omega_{m,n}|^2\geq2R}\left|\frac{\xi}{\omega_{m,n}(\omega_{m,n}^2+\xi)}\right|\leq2R\sum_{|\omega_{m,n}|\geq\sqrt{2R}}|\omega_{m,n}|^{-3}=O_\ell(\sqrt R).
\]
For the middle series with the sum taken over $\frac{R}{2}<|\omega_{m,n}|^2<2R$, there are $O_\ell(R)$ terms and each is bounded by $O_\ell(\frac{\sqrt R}{d_R})$, provided that for any $|\xi|=R$ in a set $V\subset\mathbb{C}$, 
\begin{equation}{\label{eqn:assumption}}
|\xi+\omega_{m,n}^2|\geq d_R>0
\text{ for all }-\omega_{m,n}^2\in\Omega_\ell.
\end{equation}
Then, 
\begin{equation}\label{eq:M-growth}
|M_\ell(\xi)|\leq C_\ell\left(1+\sqrt{R}+\frac{R^{\frac{3}{2}}}{d_R}\right),\text{ and }
|\widehat{U}_\ell(\xi)|\leq C_\ell\left(1+\frac{R}{d_R}\right)
\text{ for any } \xi\in V\text{ and }|\xi|=R,
\end{equation}
if the set $V$ satisfies the assumption (\ref{eqn:assumption}). 

Now we estimate $|\widehat{U}_\ell(\xi)|$ for the cases where $V$ is a ray and an arc, respectively.

(i) Since $\Omega_\ell$ is countable, we can take an order for $\{\rho_k\}_{k\geq1}$ with $|\rho_k|\leq|\rho_{k+1}|$ for all $k\geq 1$, and denote the set of directions close to the singular point $\rho_k$ by
\[E_k=\left\{\alpha\in(0,2\pi):d(\alpha,\arg\rho_k)<|\rho_k|^{-3}\right\},\]
Since $\sum_k|E_k|\leq 2\sum_{\rho_k\in\Omega_\ell,\ |\rho|\geq1}|\rho_k|^{-3}<\infty$, we have
\[\left|\bigcap_{N=1}^{\infty}\bigcup_{k=N}^{\infty}E_k\right|\leq\lim_{N\to\infty}\sum_{k=N}^{\infty}|E_k|=0.
\]
Thus, for almost every direction $\alpha\in(0,2\pi)$, there exists $N>0$, such that for all $k\geq N$, $d(\alpha,\arg\rho_k)\ge|\rho_k|^{-3}$. Fix such an $\alpha$ and denote $\xi=re^{\ii\alpha}$.
$$|\xi-\rho_k|^2
=(|\xi|-|\rho_k|)^2+4|\xi|\left|\rho_k\right|\sin^2\frac{d(\alpha,\arg\rho_k)}{2}
\geq 4|\xi|\left|\rho_k\right|\frac{d^2(\alpha,\arg\rho_k)}{\pi^2}
\geq \frac{4|\xi|}{\pi^2}\frac{1}{|\rho_k|^5}.$$
This gives $|\xi-\rho_k|\geq\frac{1}{\sqrt{8}\pi r^2}$ for all singular points satisfying $\frac{r}{2}<|\rho_k|<2r$. Taking $d_r=\frac{1}{\sqrt{8}\pi}r^{-2}$ in \eqref{eq:M-growth}, we obtain
\[
M_\ell(re^{\ii\alpha})=O_{\ell,\alpha}(r^{\frac{7}{2}}),\text{ and }\widehat U_\ell(re^{\ii\alpha})=O_{\ell,\alpha}(r^3),
\]
as $r\to \infty$ along the ray $\{re^{\ii\alpha}:r>0\}$. By the Lebesgue dominated convergence theorem, the Laplace integral converges absolutely for $\Re(\tau e^{\ii\alpha})>0$.

(ii) For sufficiently large $N$, $\#\{N-1\leq|-\omega_{m,n}^2|\leq N+2\}=O_\ell(N)$. Thus, there exists $\eta>0$, and $R_N\in[N,N+1]$ such that for any $|\xi|=R_N$ and $-\omega_{m,n}^2\in\Omega_\ell$,
$$\left|\xi-(-\omega_{m,n}^2)\right|\geq \left|R_N-|\omega_{m,n}^2|\right|\geq\frac{\eta}{N}\geq\frac{\eta}{R_N}.
$$
Taking $d_{R_N}=\frac{\eta}{R_N}$ in \eqref{eq:M-growth}, we obtain
\[
\sup_{|\xi|=R_N}|M_\ell(\xi)|=O_\ell(R_N^{\frac{5}{2}}),\text{ and }\sup_{|\xi|=R_N}|\widehat U_\ell(\xi)|=O_\ell(R_N^2), \text{ as } N\to\infty.
\]
\end{proof}

Now, we consider the Stokes phenomenon of the single-orbit series.
\begin{theorem}\label{thm:stokes}
Choose $\sigma\in\{1,-1\}$ and fix the branch determined by $\sqrt{re^{\ii\alpha}}=\sigma\sqrt{r}e^{\ii\alpha/2}$. Let $\theta$ be a singular direction, and $\{\theta\pm \delta_k\}_{k\geq 0}$ be any sequence of non-singular directions on the chosen branch such that the corresponding Laplace integrals $\mathcal L_{\theta\pm\delta_k}$ converge for every $k\in\mathbb{Z}_{\geq 0}$ and $\delta_k\to 0^+$ as $k\to\infty$. 

Then
\begin{equation}\label{eqn:stokes of single-orbit}
\lim_{k\to\infty}\left(\mathcal{L}_{\theta-\delta_k}\widehat{U}_\ell(\tau)-\mathcal{L}_{\theta+\delta_k}\widehat{U}_\ell(\tau)\right)
=\sigma\sum_{j=0}^{\infty}e^{\tau(2j+1)^2\omega_{m_0,n_0}^2}, \text{ for all }\tau \text{ with }\Re \tau e^{\ii \theta}>0,
\end{equation}
where $-\omega_{m_0,n_0}^2$ is the pole of smallest modulus on the ray $\arg\xi=\theta$.
\end{theorem}
\begin{proof}
For any fixed compact subset $K\subset \{\tau\in\mathbb{C}:\Re \tau e^{\ii \theta}>0\}$, there exist $c_K>0$ and $\delta_K>0$, such that for any $|\phi-\theta|<\delta_K$,
\[
K\subset\bigcap_{\theta-\delta_K<\phi<\theta+\delta_K}\{\Re(\tau e^{i\phi})\geq c_K\}.
\]
Choose $\delta_K$ sufficiently small in $(0,2\pi)$, and consider the directions $\theta\pm\delta_k\in(\theta-\delta_K,\theta+\delta_K)$. By Lemma \ref{lem:growth}, there exists $\{R_N\}_{N\in\mathbb{Z}_{\geq 0}}$ such that for any $\tau\in K$, there is
$$
\begin{aligned}
\mathcal{L}_{\theta-\delta_k}\widehat{U}_\ell(\tau)-\mathcal{L}_{\theta+\delta_k}\widehat{U}_\ell(\tau)
&=\lim_{\substack{\varepsilon\to 0\\R_N\to\infty}}\int_{\varepsilon e^{\ii(\theta-\delta_k)}}^{R_Ne^{\ii(\theta-\delta_k)}}\widehat{U}_\ell(\xi) e^{-\tau\xi}\dd\xi-\lim_{\substack{\varepsilon\to 0\\R_N\to\infty}}\int_{\varepsilon e^{\ii(\theta+\delta_k)}}^{R_Ne^{\ii(\theta+\delta_k)}}\widehat {U}_\ell(\xi) e^{-\tau\xi}\dd\xi\\
&=\lim_{\varepsilon\to 0}\int_{\varepsilon e^{\ii(\theta-\delta_k)}}^{\varepsilon e^{\ii(\theta+\delta_k)}}\widehat{U}_\ell(\xi) e^{-\tau\xi}\dd\xi
+\lim_{R_N\to\infty}\int_{R_N e^{\ii(\theta+\delta_k)}}^{R_N e^{\ii(\theta-\delta_k)}}\widehat{U}_\ell(\xi) e^{-\tau\xi}\dd\xi\\
&+\sigma\lim_{\substack{\varepsilon\to 0\\R_N\to\infty}}\sum_{\substack{\theta-\delta_k<\arg(-\omega_{m,n}^2)<\theta+\delta_k\\ \varepsilon<\left|\omega_{m,n}^2\right|<R_N}}e^{\tau\omega_{m,n}^2},
\end{aligned}
$$
where the third term follows from the residue theorem and (\ref{eq:local-ML}), which give $2\pi\ii\Res\left(\widehat{U}_\ell(\xi);\xi=-\omega_{m,n}^2\right)=\sigma$ on the fixed branch. Since $\widehat{U}_\ell(\xi)=O_\ell(|\xi|^{-\frac{1}{2}})$ as $\xi\to0$ and $\sup_{|\xi|=R_N}|\widehat U_\ell(\xi)|=O_\ell(R_N^2)$ as $R_N\to\infty$, the integrals over the arcs are respectively $O(\varepsilon^{\frac{1}{2}})$ and $O(R_N^3e^{-c_KR_N})$. Thus, for any $\theta\pm\delta_k\in(\theta-\delta_K,\theta+\delta_K)$, 
\[\mathcal{L}_{\theta-\delta_k}\widehat{U}_\ell(\tau)-\mathcal{L}_{\theta+\delta_k}\widehat{U}_\ell(\tau)=\sigma\sum_{\theta-\delta_k<\arg(-\omega_{m,n}^2)<\theta+\delta_k}e^{\tau\omega_{m,n}^2},\text{ for all }\tau\in K.\]
Denote by $-\omega_{m_0,n_0}^2$ the pole of smallest modulus on the ray $\arg\xi=\theta$, and consider a pole $-\omega_{m,n}^2$ on another ray, i.e. $0<|\arg(-\omega_{m,n}^2)-\theta|<\delta_k<\delta_K$. Then 
$|\Im(\overline{\omega_{m_0,n_0}}\omega_{m,n})|=|\omega_{m_0,n_0}||\omega_{m,n}|\,|\sin(\arg \omega_{m,n}-\arg\omega_{m_0,n_0})|<\frac{\delta_k}{2}|\omega_{m_0,n_0}||\omega_{m,n}|$; on the other hand $\left|\Im(\overline{\omega_{m_0,n_0}}\omega_{m,n})\right|=\frac{\pi}{\ell}\left|(2m_0+1)n-(2m+1)n_0\right|\geq\frac{\pi}{\ell}$. Thus
\[|\omega_{m,n}|>\frac{2\pi}{\ell|\omega_{m_0,n_0}|\delta_k},\]
for every singular point in $\{-\omega_{m,n}^2\in\Omega_\ell: \theta-\delta_k<\arg(-\omega_{m,n}^2)<\theta+\delta_k\}$. At the same time, from $\delta_k<\delta_K$, we have $|e^{\tau\omega_{m,n}^2}|\leq e^{-c_K|\omega_{m,n}|^2}$ for all $\tau\in K$. Then,
\[
\sum_{\substack{0<|\arg(-\omega_{m,n}^2)-\theta|<\delta_k}}|e^{\tau\omega_{m,n}^2}|
\leq e^{-c_K'/\delta_k^2}\sum_{m\geq0,\ n\in\mathbb{Z}}e^{-c_K|\omega_{m,n}|^2/2}
=O(e^{-c_K'/\delta_k^2}),\quad \forall \tau\in K,
\]
with $c_K'=\frac{2c_K\pi^2}{(\ell|\omega_{m_0,n_0}|)^2}>0$. Letting $\delta_k\to0^+$, and using the form of singular points in Lemma \ref{lemma:singular direction}(ii), we obtain the equation (\ref{eqn:stokes of single-orbit}).
\end{proof}
\begin{remark}
By the bijection between Borel singularities and cylinder resonances established in Proposition \ref{pro:cylinder},
 Theorem \ref{thm:stokes} implies that the Stokes phenomenon of single-orbit series is determined by the cylinder resonance spectra.
\end{remark}

\section{Conclusions}\label{sec:conclusions}
We have obtained the $1-$Gevrey asymptotic expansion of the single-orbit term. We call this expansion the single-orbit series, compute its Borel transform, and find that the singularities in the Borel plane correspond to the associated hyperbolic cylinder resonance spectra. By analyzing the growth order and distribution of the singular points of $\widehat{U}_\ell(\xi)$, we give the limits of the differences between the Laplace integrals along different directions, and the resulting Stokes structure can be represented by a sum of exponential terms associated with the resonance spectra.

 This framework answers the first question that how spectral information appears in the asymptotic expansion of an orbit term. It gives a new viewpoint to understand the relation between periodic orbits in classical mechanics and the spectra in quantum mechanics: As a local example, fix a primitive orbit length $\ell$ on $X$, the resonance spectra of a hyperbolic cylinder obtained by $\mathbb{H}/\{\ell\sim e^{\ell}\}$ emerge as the non-zero singular points of the Borel transform of the long-time asymptotic expansion of the fixed single-orbit term. Furthermore, the spectra on hyperbolic cylinders govern the Stokes structure of the single-orbit series as shown in Theorem \ref{thm:stokes}.

However, the answer to the second question if a single-orbit series can give a spectrum on $X$ is still unclear in this framework at the moment. The Borel transform of a finite sum of single-orbit series only give the spectrum information of the associated hyperbolic cylinders, rather than the global spectrum on $X$. It seems that the global spectrum can only be visible in the Borel transform of the summation over ALL primitive single-orbit series. By summing more and more different single-orbit series, until all primitive periodic orbits have been included, we hope to see how the singularities of the corresponding Borel transforms evolve throughout this process. This may reveal how the local spectral data associated with the hyperbolic cylinders gradually become the global spectrum as the orbit information in classical mechanics increases. These will be studied in our future work.

\section*{Acknowledgments}
{The author expresses sincere gratitude to Professor Shanzhong Sun for suggesting this problem, insightful discussions, and careful reading of the manuscript. His guidance was essential to the completion of this work. }

\section*{Statement on AI assistance}
The mathematical results and manuscript were independently developed and written by the author. During the revision process, ChatGPT 6 (OpenAI) was consulted for feedback on English expression and mathematical arguments. The author is responsible for the verification and final form of all mathematical arguments and text in this paper.

\end{document}